\documentstyle[aas2pp4,flushrt]{article}
\def\etal{{et al.}}

\leftline{Accepted by ApJ Letters April 2000}
\begin{document}

\title{The Missing Link: Early Methane (``T'') Dwarfs \\
in the Sloan Digital Sky Survey}

\author{S.K. Leggett\altaffilmark{\ref{UKIRT}},
T.R. Geballe\altaffilmark{\ref{Gemini}},
Xiaohui Fan\altaffilmark{\ref{Princeton}},
Donald P. Schneider\altaffilmark{\ref{PennState}}, 
James E. Gunn\altaffilmark{\ref{Princeton}}, 
Robert H. Lupton\altaffilmark{\ref{Princeton}}, 
G. R. Knapp\altaffilmark{\ref{Princeton}},
Michael A. Strauss\altaffilmark{\ref{Princeton}},
Alex McDaniel\altaffilmark{\ref{Princeton}},
David Golimowski\altaffilmark{\ref{JHU}},
Todd Henry\altaffilmark{\ref{JHU}},
Eric Peng\altaffilmark{\ref{JHU}},
Zlatan I. Tsvetanov\altaffilmark{\ref{JHU}},
Alan Uomoto\altaffilmark{\ref{JHU}},
Wei Zheng\altaffilmark{\ref{JHU}}
G.J. Hill\altaffilmark{\ref{Texas}},
L.W. Ramsey\altaffilmark{\ref{PennState}},
Scott F. Anderson\altaffilmark{\ref{Washington}},
James A. Annis\altaffilmark{\ref{Fermilab}},
Neta A. Bahcall\altaffilmark{\ref{Princeton}},
J. Brinkmann\altaffilmark{\ref{APO}}, 
Bing Chen\altaffilmark{\ref{JHU}},
Istv\'an Csabai\altaffilmark{\ref{JHU},\ref{Eotvos}},
Masataka Fukugita\altaffilmark{\ref{CosmicRay},\ref{IAS}},
G. S. Hennessy\altaffilmark{\ref{USNO}},
Robert B. Hindsley\altaffilmark{\ref{USNO}},
\v{Z}eljko Ivezi\'{c}\altaffilmark{\ref{Princeton}},
D.Q. Lamb\altaffilmark{\ref{Chicago}},
Jeffrey A. Munn\altaffilmark{\ref{Flagstaff}},
Jeffrey R. Pier\altaffilmark{\ref{Flagstaff}},
David J. Schlegel\altaffilmark{\ref{Princeton}},
J. Allyn Smith\altaffilmark{\ref{Michigan}},
Chris Stoughton\altaffilmark{\ref{Fermilab}},
A. R. Thakar\altaffilmark{\ref{JHU}}, and
Donald G. York\altaffilmark{\ref{Chicago}}
}

\newcounter{address}
\setcounter{address}{1}
\altaffiltext{\theaddress}{United Kingdom Infrared Telescope, Joint Astronomy
Centre, 660 North A'ohoku Place, Hilo, Hawaii 96720
\label{UKIRT}}
\addtocounter{address}{1}
\altaffiltext{\theaddress}{Gemini Observatory, 670 North A'ohoku Place,
Hilo, HI 96720
\label{Gemini}}
\addtocounter{address}{1}
\altaffiltext{\theaddress}{Princeton University Observatory, Princeton, NJ 08544
\label{Princeton}}
\addtocounter{address}{1}
\altaffiltext{\theaddress}{Department of Astronomy and Astrophysics,
The Pennsylvania State University,
University Park, PA 16802
\label{PennState}}
\addtocounter{address}{1}
\altaffiltext{\theaddress}{
Department of Physics and Astronomy, The Johns Hopkins University,
   3701 San Martin Drive, Baltimore, MD 21218, USA
\label{JHU}}
\addtocounter{address}{1}
\altaffiltext{\theaddress}{Department of Astronomy, McDonald Observatory,
   University of Texas, Austin, TX~78712.
\label{Texas}}
\addtocounter{address}{1}
\altaffiltext{\theaddress}{University of Washington, Department of Astronomy,
Box 351580, Seattle, WA 98195
\label{Washington}}
\addtocounter{address}{1}
\altaffiltext{\theaddress}{Fermi National Accelerator Laboratory, P.O. Box 500,
Batavia, IL 60510
\label{Fermilab}}
\addtocounter{address}{1}
\altaffiltext{\theaddress}{Apache Point Observatory, P.O. Box 59,
Sunspot, NM 88349-0059
\label{APO}}
\addtocounter{address}{1}
\altaffiltext{\theaddress}{Department of Physics of Complex Systems,
E\"otv\"os University,
   P\'azm\'any P\'eter s\'et\'any 1/A, Budapest, H-1117, Hungary
\label{Eotvos}}
\addtocounter{address}{1}
\altaffiltext{\theaddress}{Institute for Cosmic Ray Research, University of
Tokyo, Midori, Tanashi, Tokyo 188-8502, Japan
\label{CosmicRay}}
\addtocounter{address}{1}
\altaffiltext{\theaddress}{Institute for Advanced Study, Olden Lane,
Princeton, NJ 08540
\label{IAS}}
\addtocounter{address}{1}
\altaffiltext{\theaddress}{U.S. Naval Observatory, 
3450 Massachusetts Ave., NW, 
Washington, DC  20392-5420
\label{USNO}}
\addtocounter{address}{1}
\altaffiltext{\theaddress}{University of Chicago, Astronomy \& Astrophysics
Center, 5640 S. Ellis Ave., Chicago, IL 60637
\label{Chicago}}
\addtocounter{address}{1}
\altaffiltext{\theaddress}{U.S. Naval Observatory, Flagstaff Station, 
P.O. Box 1149, 
Flagstaff, AZ  86002-1149
\label{Flagstaff}}
\addtocounter{address}{1}
\altaffiltext{\theaddress}{University of Michigan, Department of Physics,
500 East University, Ann Arbor, MI 48109
\label{Michigan}}

\begin{abstract}

We report the discovery of three cool brown dwarfs which fall in the
effective temperature gap between the latest L dwarfs currently known,
with no methane absorption bands in the 1---2.5~$\mu$m range, and the 
previously known methane (T)
dwarfs, whose spectra are dominated by methane and water. The newly
discovered objects were detected as very red objects in the Sloan Digital
Sky Survey imaging data and have JHK colors between the red L
dwarfs and the blue Gl~229B--like T dwarfs. They show both CO and CH$_4$
absorption in their near--infrared spectra in addition to H$_2$O, with
weaker CH$_4$ absorption features in the H and K bands than those in all 
other methane dwarfs reported to date. Due to the presence of CH$_4$ in these
bands, we propose that these objects are  early T dwarfs.  The 
three  form part of the brown dwarf spectral sequence and fill in the large gap
in the overall spectral sequence from the hottest main sequence stars to
the coolest methane dwarfs currently known.

\end{abstract}
\keywords{brown dwarfs; surveys}

\section{Introduction}

Brown dwarfs, gravitationally condensed objects whose masses are too low
for equilibrium hydrogen burning, occupy the mass range between the lowest
mass stars ($\rm \sim 0.07 ~ M_{\odot}$; Burrows et al. 1997) and the
giant extrasolar planets ($\rm \sim 0.01 ~ M_{\odot}$; Marcy \& Butler
1998).  The lowest mass stars and brown dwarfs only slightly cooler than M
dwarfs are classified as spectral type ``L'' (Kirkpatrick et al. 1999,
Mart\'{\i}n et al. 1997) and have T$_{eff}$ in the range
$\sim$1500---2000~K.  Determining if
L--type objects are brown dwarfs is difficult, because the luminosities and
effective temperatures of brown dwarfs are a function of both age and mass
(e.g. Burrows et al. 1997). The first unambiguous brown dwarf, Gl~229B,
was discovered as a companion to a nearby M dwarf by Nakajima et al.
(1995). This object is cooler than any L dwarf and
has CH$_4$ in its atmosphere implying $\rm T_{eff}<$1300 K (e.g. 
Fegley \& Lodders 1996) and a sub--stellar nature. The recent discoveries of 
methane dwarfs (tentatively given
a ``T'' spectral clasification) in the field by Strauss et al. (1999),
Burgasser et al. (1999, 2000), Cuby et al. (1999) and Tsvetanov et al.
(2000) demonstrate that objects like Gl~229B can form singly.

A striking characteristic of known T dwarfs is the similarity
of their spectra, which resemble those of the gaseous solar system planets and
are very different from even the coolest L dwarfs. The L dwarf spectra are
characterized by the disappearance of the TiO and VO bands (which are strong 
in M dwarf spectra), the presence of
atomic alkali lines and CO bands, and increasing depth of the $\rm H_2O$
bands as T$_{eff}$ decreases.  The T dwarfs are characterized by very
deep $\rm H_2O$ and $\rm CH_4$ bands in the 1--3 $\rm \mu m$ region.
Alkali metal features are present 
between the bands, but the 2.3$\rm \mu m$ CO bands are absent, due to 
reduced CO abundance and the overwhelming strength of the CH$_4$
absorption.

A large remaining gap in the stellar to gas giant planet
spectral sequence lies between the previously observed types L and T.  In
this paper, we report photometry and 0.8---2.5 $\rm \mu m$ spectroscopy of
three very red, faint objects identified in the imaging data of the Sloan
Digital Sky Survey (SDSS) and observed at the United Kingdom Infrared
Telescope (UKIRT) and the Hobby-Eberly Telescope (HET). These objects have
spectra intermediate between those of the previously known L and T dwarfs,
showing the onset of CH$_4$ absorption at 1.6 and 2.2 $\rm \mu m$, while
still retaining observable CO absorption at 2.3 $\rm \mu m$.  We propose
that they represent the warm end of the ``T'' spectral sequence, yet to be
defined in detail.

\begin{deluxetable}{rrrrrrrrrr}
\footnotesize
\tablenum{1}
\tablecaption{Coordinates and Photometry}
\tablehead{
\colhead{RA} & \colhead{Dec} & \colhead{$i^*$} &
\colhead{$z^*$}& \colhead{J}& \colhead{H} & \colhead{K}& \colhead{J} & \colhead{J$-$H}  & \colhead{H$-$K} \nl
\multicolumn{2}{c}{(2000)} & \multicolumn{2}{c}{(AB)} & \multicolumn{3}{c}{(UKIRT--UFTI)} & \multicolumn{3}{c}{(UKIRT--IRCAM)}\nl
}
\tablecolumns{7}
\startdata 
05 39 51.99 & $-$00 59 02.0 & 19.04 $\pm$0.02\phm{)}& 16.73 $\pm$0.01 
& 13.85 $\pm$0.03 & 13.04 $\pm$0.03 & 12.40 $\pm$0.03 & 13.94  & 0.97  & 0.53  \nl
08 37 17.21 &$-$00 00 18.0 &(23.51 $\pm$0.42) & 19.95 $\pm$0.09  
& 16.90 $\pm$0.05 & 16.21 $\pm$0.05 & 15.98 $\pm$0.05 & 17.08  & 0.93  & 0.11  \nl
10 21 09.69 & $-$03 04 20.1 &  (23.73 $\pm$0.58)   & 19.28 $\pm$0.05   
& 15.88 $\pm$0.03 & 15.41 $\pm$0.03 & 15.26 $\pm$0.05 & 16.12  & 0.74 & 0.08   \nl
12 54 53.90 &$-$01 22 47.4 &  22.22 $\pm$0.28\phm{)} & 18.00 $\pm$0.04 
& 14.66 $\pm$0.03 & 14.13 $\pm$0.03 & 13.84 $\pm$0.03 & 14.90 & 0.84  & 0.15 \nl
\enddata
\tablecomments{ $i^*$ and $z^*$ are asinh magnitudes on the
AB system (Lupton et al. 1999).  Zero flux corresponds to $i^*=23.89$ and
$z^*=22.47$.}
\end{deluxetable}

\section{Observations}

\subsection{SDSS Photometry and Object Selection}

The three objects described here were selected from SDSS photometric data.  
SDSS photometry is obtained with a CCD camera at Apache Point Observatory 
(APO) which images the sky almost
simultaneously in five filters: $u'$, $g'$, $r'$, $i'$, and $z'$ (the data
presented here use a preliminary calibration; while we denote the bands by
$u'$ etc, the magnitudes are denoted by $u^*$ etc.).  The details of the data
acquisition and the photometric and astrometric calibration are described
by Gunn et al. (1998), York et al. (2000) and Lupton et al. (2000).  SDSS
photometry is in the $AB_{\nu}$ system (Fukugita et al. 1996) and the
magnitude scale is modified to deal with low signal--to--noise ratios (Lupton
et al. 1999).

The L and T dwarfs identified in the SDSS (e.g. Strauss et al. 1999, Fan et al.
2000) are undetected in the $u'$ and $g'$ bands. The L dwarfs have
$i^*-z^* >$ 1.6 while the T dwarfs have $i^*-z^* >$ 3.  We prepared a
candidate L and T dwarf list for follow--up observations by
searching the SDSS data for point sources with $i^*-z^* >$ 1.6. 
Because of their extremely red colors, some very red, faint objects are
detected only in
the $z'$ band and a search for them can be contaminated by defects.  
Therefore, additional constraints were imposed: (1) all objects blended with
neighbors, or affected by data defects, were removed; (2) an object was
required to be detected at $\geq$3$\sigma$ twice: in both $i'$ and $z'$,
in two observations of that region of sky, or detected in the 2MASS
database; (3) a $z'$--band--only single detection was required to 
have $z'<$19.0.  The three objects discussed in this paper were selected from
a total sky area of 225~deg$^2$, giving an approximate and very
preliminary surface density similar to 
the 1 per 75~deg$^2$ for SDSS T dwarfs found by Tsvetanov et al. (2000).

Table~1 gives, for the three new objects, the J2000
position (accurate to $0\farcs 2$) and AB magnitudes at
$i'$ and $z'$.    Table~1 also lists photometry
obtained for the L dwarf SDSS~0539 (Fan et al. 2000).  Hereafter 
we identify the objects by the first four digits of their Right Ascension.
SDSS~0837 was found in two SDSS runs, 1999 March 21 and
2000 February 8;  the $z^*$ values agree to 0.1~mag and the positions
to $0\farcs 3$.  SDSS~1021 is a $z'$--only detection  in
data from 2000 February 12 and is also in the 2MASS database.  
The brightest object, SDSS~1254, has
$z^*=$18.0 and was found in data taken on 2000 February 2.  Finding
charts for the new objects are shown in Figure 1.

\begin{figure}[!!h]
\figurenum{1}
\plotone{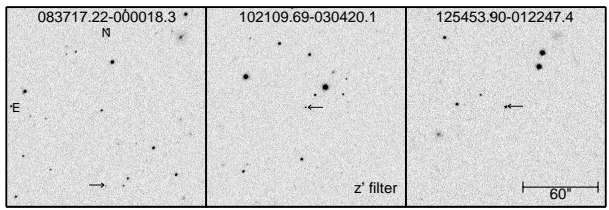}
\caption{Finding charts for the new early T dwarfs.}
\end{figure}

\subsection{UKIRT Photometry}

J, H and K photometry was obtained using UKIRT's near--infrared camera, UFTI,
for SDSS 0837 and SDSS 1254 on 2000 March 2 and for SDSS 0539 and SDSS 1021 
on March 14.  Both nights were photometric with seeing $\sim$0\farcs8.  The 
results are given in Table~1 in the UKIRT (not AB) system --- we give JHK
on the UFTI system as well as colors on the established IRCAM system for
comparison to earlier work.    Figure~2
shows a composite color--color plot containing the data from Table 1 and,
for comparison, a sample of late M and L dwarfs as well as the SDSS T
dwarfs SDSS 1624 and SDSS 1346 (Strauss et al. 1999; Tsvetanov et al.
2000), whose spectra closely resemble that of Gl 229B.  Lower limits are
shown for $i^* - z^*$ for the $z'$--only detections.
The late M and L
dwarfs show a steady progression towards redder J$-$K and $i^*-z^*$ colors
with later spectral type, however the T dwarfs have redder $i^*-z^*$
colors and {\it bluer} J$-$K colors due to the strong absorption by CH$_4$
in the H and K bands.  The J$-$K colors of the new objects lie between
those of the previously known L and T dwarfs.

\begin{figure}[!!h]
\figurenum{2}
\epsscale{0.80}
\plotone{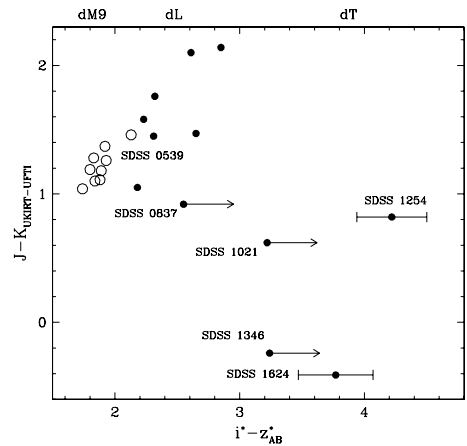}
\caption{ SDSS $i^*-z^*$ against UKIRT--UFTI J$-$K;
open symbols have J$-$K from 2MASS. 
Errors in $i^*-z^*$  are $\leq \pm 0.15$~mag except where indicated; 
measurement error and system
differences in the 2MASS J$-$K are $\sim$0.2~mag; error in the 
UKIRT J$-$K is 0.05~mag.}
\end{figure}

\subsection{HET Spectroscopy}

A red spectrum of SDSS 1254 was obtained on 2000 February 29 with the Low
Resolution Spectrograph (LRS; Hill et al. 1998a,b, Schneider et al. 2000) 
at the prime focus of
the of the HET (Ramsey et al. 1998). The spectrum covers the wavelength
range 5100--9800~\AA\ at a resolution of~$\sim$~20~\AA, but no flux is
detected below $\sim$6500~\AA.

\subsection{UKIRT Spectroscopy}

Spectra of the objects in Table 1 were obtained in the J, H and K
bands at UKIRT on 2000 February 28 --- March 1 and March 13---15
using the facility spectrograph CGS4 (Mountain et al. 1990) at
R$\sim$400.  Integration times were 30---60 minutes per band.
Telluric absorption was removed and the broad--band spectral shapes
corrected by comparison with spectra of bright F stars (with 
photospheric lines removed) taken immediately before and/or after each
observation.  The spectra in each band were scaled to match the
UKIRT magnitudes using the UFTI filter profiles.  UKIRT z--band spectra 
were also obtained for SDSS 1254 and SDSS 0539.  For   SDSS 1254
the HET spectrum is used for  $\lambda < 0.84 \mu$m, and for SDSS 0539
the red APO spectrum (Fan et al. 2000) is used for  $\lambda < 0.89 \mu$m.

\begin{figure}[!!h]
\figurenum{3}
\epsscale{1.0}
\plotone{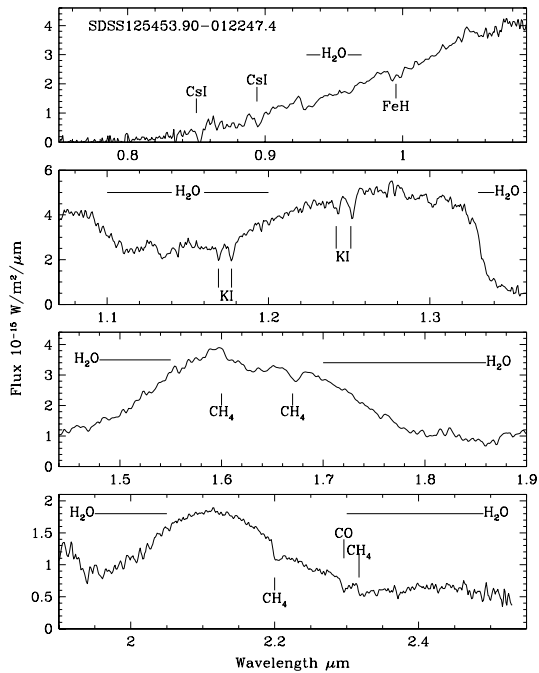}
\caption{UKIRT z, J, H and K spectra of SDSS 1254. The z spectrum (top) is 
extended to shorter wavelengths by the HET spectrum.
The data are scaled to absolute values using UKIRT photometry.
Atomic lines (Cs I, K I), narrow molecular features (FeH, CH$_4$), and
bandheads 
(CH$_4$, CO) are indicated by vertical lines while the broad H$_2$O
bands are indicated by horizontal lines.}
\end{figure}

The final flux--calibrated spectra are shown in Figures 3 and 4.  Figure 3
shows the spectrum of SDSS 1254 split into four panels, with the main
absorption features marked.  Figure 4 compares the spectra of the 
new objects with those of the L5 dwarf SDSS 0539 (Fan et al. 2000)
 and the T dwarf SDSS
1624 (Strauss et al. 1999).  The spectra show: increasingly red colors at
$\lambda <1~\mu$m, probably due to pressure broadened Na I and K I 
absorption (Burrows et al. 2000) combined with decreasing T$_{eff}$; 
increasing absorption in the H$_2$O bands at
1.15~$\mu$m, 1.4~$\mu$m and 1.9~$\mu$m; increasing absorption at
1.6---1.7~$\mu$m and longward of 2.2~$\mu$m by CH$_4$ combination and
overtone bands; decreasing absorption in the CO 2.3$\mu m$ band. In
particular, the absorption maxima of the CH$_4$ 2$\nu$$_2$+$\nu$$_3$,
2$\nu$$_2$, and $\nu$$_2$+$\nu$$_3$ bands at 1.63, 1.67, and 2.20~$\mu$m
become progressively deeper. The CO 2--0 bandhead at 2.294~$\mu$m is
seen in the new objects, but the 3--1 bandhead at 2.323~$\mu$m is
absent; it is considerably weaker at these temperatures and also is
overwhelmed by the strong CH$_4 ~\nu _3 + \nu _4$ absorption at
2.315~$\mu$m.  

\begin{figure}[!!h]
\figurenum{4}
\epsscale{1.0}
\plotone{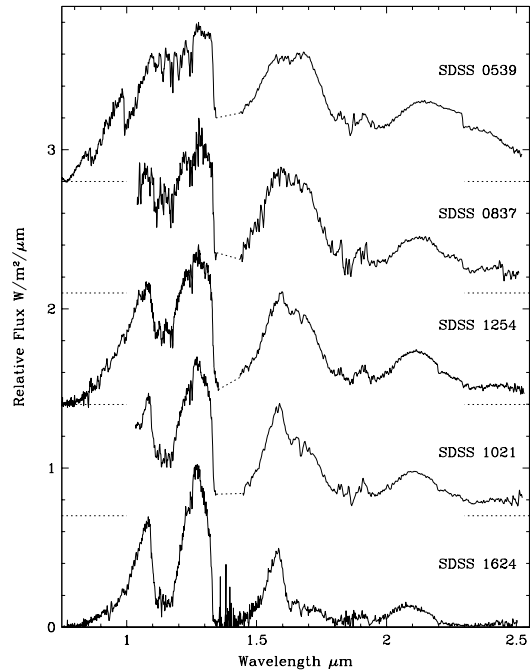}
\caption{A T dwarf spectral sequence.  The spectra
of the three early T dwarfs (SDSS 0837, SDSS 1254 and SDSS 1021) are 
ordered by increasing CH$_4$ absorption.   Also shown are  the L
dwarf SDSS 0539 ($\sim$L5, Fan et al. 2000, optical data from Fan et al.)
and the later T dwarf SDSS 1624 (Strauss et al. 1999).
No data were obtained near 1.4$\mu$m due to telluric H$_2$O
absorption. The spectra are scaled to the flux peak at 1.27$\mu$m
and offset; dotted lines indicate zero levels.}
\end{figure}

\section{Discussion and Conclusions}

The spectra in Figure 4 reveal a clear spectral sequence.  Thus SDSS 0837,
SDSS 1254 and SDSS 1021 are examples of the sought after ``L/T
transition'' objects. However, as suggested by Figure 4, they might more
properly be regarded as examples of early T dwarfs, where the
``T'' spectral type is defined by the presence of CH$_4$ absorption in the
H and K bands. The strongest short wavelength infrared band of CH$_4$ is
the $\nu$$_2$ band, centered at 3.3~$\mu$m, and objects in which this band
is present but the shorter wavelength  bands reported here are absent
must exist over a narrow temperature range. However, measurements at
3.3~$\mu$m are considerably more difficult from the ground than are H and
K band measurements, and objects which show only the 3.3~$\mu$m CH$_4$
band perhaps should be classified as L rather than T (Kirkpatrick et al.
(1999) suggest that the T class be defined by the presence of CH$_4$
absorption
in the K band).  As the three objects
presented here are at the CO/CH$_4$ transition temperature they are likely to 
have
T$_{eff} \approx 1300$~K  (Fegley \& Lodders 1996)  and therefore masses of
20---70~M$_{Jupiter}$ for ages in the range 0.3---5~Gyr (Burrows et al. 1997).

Figure 2 shows that SDSS easily
identifies T dwarfs by their extremely red $i^*-z^*$ color, including the
early T dwarfs identified in the present paper.  On the other hand, the
J$-$K colors are similar to those of the common M dwarfs and these early T
objects are thus very difficult to select on the basis of the
near--infrared colors alone (although the later T dwarfs are blue and
{\it can} be selected this way).  Many more L and T dwarfs will be found
in the SDSS imaging data.

\footnotesize
\acknowledgements The SDSS is a joint project of the University of
Chicago, Fermilab, the Institute for Advanced Study, the Japan
Participation Group, Johns Hopkins University, Max--Planck--Institute for
Astronomy, Princeton University, United States Naval Observatory, and the
University of Washington.  The site, Apache Point Observatory, is operated
by the Astrophysical Research Consortium.  Funding has been provided by
the Alfred P. Sloan Foundation, the member institutions, NASA, NSF, the
U.S. DoE, and Monbusho Japan. The SDSS Web site is at
{\tt http://www.sdss.org/}. UKIRT is operated by the Joint Astronomy
Centre on behalf of the UK Particle Physics and Astronomy Research
Council.  We thank Tim Carroll for his expert and cheerful help with the
UKIRT observations, and M.-C. Liang for assistance during the 28 Feb
-- March 1 run. The Hobby-Eberly Telescope is a joint project of the
University of Texas at Austin,  Pennsylvania State University, Stanford
University, Ludwig--Maximillians--Universit\"at M\"unchen, and
Georg--August--Universit\"at G\"ottingen.  The HET principal benefactors
are William P. Hobby and Robert E. Eberly.  We thank Phillip MacQueen,
Grant Hill, Matthew Sheltrone, and Marsha Wolf for help
with the HET data.  XF and MAS acknowledge support from the Research
Corporation, NSF grant AST96-16901, the Princeton University Research
Board, and a Porter O. Jacobus Fellowship. GRK is grateful for support
from Princeton University and from NASA via grant NAG-6734.  DPS thanks
the NSF for support via grant AST99-00703.






\small
\begin{references}
\reference{}
Burgasser, A.J. \etal\ 1999, ApJ, 522, L65 
\reference{}
Burgasser, A.J. \etal\ 2000, AJ (in press: astro-ph/0004239)
\reference{}
Burrows, A.,  Marley, M.S., \& Sharp, C.M., 2000, ApJ, 531, 438 
\reference{}
Burrows, A. \etal\ 1997, ApJ, 491, 856 
\reference{}
Cuby, J.G., Saracco, P., Moorwood, A.F.M., D'Odorico, S., 
 Lidman, C., Comer\'on, F., \& Spyromilio, J. 1999, A\&A, 349, L41
\reference{}
Fan, X.  \etal\ 2000, AJ, 119, 928 
\reference{}
Fegley, B. \& Lodders, K., 1996, ApJ, 472, L37 
\reference{}
Fukugita, M., Ichikawa, T., Gunn, J.E.,
Doi, M., Shimasaku, K., \& Schneider, D.P. 1996, AJ, 111, 1748 
\reference{}
Gunn, J.E.  \etal\ 1998, AJ, 116, 3040 
\reference{}
Hill, G.J., Nicklas, H.E., MacQueen, P.J., Tejada, C., Cobos~Duenas,
     F.J., \& Mitsch, W. 1998a, Proc. SPIE, 3355, 375 
\reference{}
Hill, G.J., Nicklas, H.E., MacQueen, P.J., Mitsch, W., Wellem, W.,
     Altmann, W., Wesley, G.L., \& Ray, F.B. 1998b, in Proc. SPIE, 3355, 433 
\reference{}
Kirkpatrick, J.D. \etal\ 1999, ApJ, 519, 802 
\reference{}
Lupton, R.H., Gunn, J.E., \& Szalay, A. 1999, AJ, 118, 1406 
\reference{}
Lupton, R.H. \etal\ 2000, in preparation 
\reference{}
Marcy, G.W., \& Butler, R.P. 1998, in ``Brown Dwarfs and Extrasolar
Planets'', ed.~R. Rebolo, E.L. Mart\'{\i}n \& M.R. Zapatero-Osorio,
A.S.P. Conf. Ser. 134, 128 
\reference{}
Mart\'{\i}n, E.L., Basri, G., Delfosse, X., \&
 Forveille, T. 1997, A\&A, 327, L29 
\reference{}
Mountain, C.M., Robertson, D., Lee, T.J., \& Wade, R. 1990, Proc. SPIE,
  1235, 25 
\reference{}
Nakajima, T., Oppenheimer, 
B.R., Kulkarni, S.R., Golimowski, D.A., Matthews, K., \& Durrance,
S.T., 1995, Nature, 378, 463 
\reference{}
Ramsey, L.W. \etal\ 1998, Proc. SPIE, 3352, 34
\reference{}
Schneider, D.P. \etal\ 2000, PASP, 112, 6
\reference{}
Strauss, M.A.  \etal\ 1999, ApJ, 522, L61
\reference{}
Tsvetanov, Z.I.  \etal\ 2000, ApJ, 531, L61 
\reference{}
York, D.G.  \etal\ 2000,  submitted to AJ

\end{references}
\end{document}